\documentstyle[12pt]{article}
\catcode`\@=11
\newcommand{\ccite}[1]
{\@ifundefined{b@#1}{\bf ?}{\@nameuse{b@#1}}}
\begin{document}
\setcounter{page}{0}
\baselineskip = 20pt

\title{Quasiperiodic fields and  Bose-Einstein condensation}

\author{P. F. Borges,\thanks{e-mail: pborges@if.ufrj.br;} 
H. Boschi-Filho\thanks{e-mail: boschi@ctp.mit.edu;} 
and C. Farina\thanks{e-mail: farina@if.ufrj.br.}
\\ 
\small \it
Instituto de F\'\i sica, Universidade Federal do Rio de Janeiro \\ 
\small \it Cidade Universit\'aria, Ilha do Fund\~ao, Caixa
Postal 68528 \\ 
\small \it 21945-970 Rio de Janeiro, BRAZIL}

\date{}

\maketitle 
\thispagestyle{empty}

\begin{abstract} 
We construct a partition function for fields obeying a quasiperiodic
boundary condition at finite temperature, 
$\psi(0;\vec x)=e^{i\theta}\psi(\beta;\vec x)$, 
which interpolate continously that
ones corresponding to bosons and fermions and discuss the possibility 
of condensation for these fields.
\end{abstract}

\vfill


\vskip 1cm

\pagebreak


The concept of condensation was proposed by Einstein \cite{Einstein},
after the fundamental work of Bose \cite{Bose} on the statistics of
integer spin particles, and its first realization was given by London in a
model for the Helium superfluidity \cite{L ondon}. More recently, the
interest for this phenomenon has increased because of the experimental
results on atomic traps and the direct observation of Bose-Einstein
condensation (BEC) on gases \cite{exp} signalizing for a large number of
applications. 

Long ago, it was shown for massive bosons in the absence of external
fields that BEC can only occur if $N \ge 3$, where $N$ is the space
dimension. For massless bosons this condition reduces to $N \ge 2$,
although in this case the critical temperature is infinite.  These results
were first shown in a nonrelativistic context \cite{Huang} and later
generalized the relativistic case [\ccite{BKM}-\ccite{Kapusta}]. Very
recently, the extension of these results to curved space-times and the
inclusion of non-uniform magnetic fields in the relativistic context were
also done \cite{Toms,KirstenToms}. 

Here, we present a relativistic quantum field theory approach to BEC from
which we obtain nonrelativistic results as a particular limit. We start
from a partition function identified with a determinant for fields which
satisfy a quasiperiodic boundary condition, {\sl i.e.}, a condition which
interpolates continuously the periodic (bosonic) and antiperiodic
(fermionic) cases at finite temperature. This kind of field is usually
considered in lower dimensional systems as in the anyon theory in 2+1
dimensions \cite{DJT,Wilczek}. Recently, we have shown \cite{BBF} that the
present construction leads to a partition function which is equivalent to
the usual ones for anyons, at least for the second virial coefficient
\cite{Arovas}.  Further within our approach, we also obtain a general
condensation condition for fields which interpolate continuosly between
bosons and fermions and discuss the role of the dimension where the
condensation occurs.


Let us start by writting the partition function for a general system
described by a Hamiltonian $H$ with chemical potential $\mu$ as
\begin{equation}\label{Z}
{\cal Z}\equiv \exp\{-\beta\Omega\}= {\rm tr} \;e^{-\beta(H+ {\cal N}\mu)}, 
\end{equation}

\noindent where $\Omega$ is the free energy and $\cal N$ is the
conserved charge (conserved quantum number) of the system. As it is
well known, for relativistic charged massive bosonic fields one can 
express this partition function as a determinant \cite{Kapusta,Bernard},
namely,
\begin{equation}\label{bosondet}
{\cal Z}_{Bosons}=\left[\left.\det(-D^2+M^2)\right|_{P}\right]^{-1}\;, 
\end{equation}

\noindent where $D^2$ is the square of the covariant derivative,  
including the chemical potential, $D_\nu=(\partial_0 +
i\mu,\partial_i)$ and $M$ is the mass of the field.  This prescription
for the inclusion of the chemical potential $\mu$ coincides with the
one given by eq. (\ref{Z}) except for the introduction of an imaginary
part to $\cal Z$, which is not physically relevant.  So unless
otherwise specified, we are considering that the real part of $\cal Z$
has been taken. Note also that, the chemical potential appears squared
in the Klein-Gordon (KG) operator, $-D^2+M^2$.  This is a consequence
of representing the partition function in phase space and then
integrating over momenta (see \cite{Kapusta,Actor} for details).  The
label {\it P} means that the eigenvalues of this operator are
subjected to periodic boundary conditions and hence they are given by
\begin{equation}\label{eigenvalues}
\lambda_{nk}=(\omega_n + i\mu)^2+\vec k^2+M^2\;, 
\end{equation}

\noindent where $\omega_n=2n\pi/\beta$, with $n\in {\sf Z}\!\! {\sf
Z}$, are the Matsubara frequencies \cite{Matsubara} for bosonic fields
and $\vec k\, \in \, {{\sf I}\!\!\; {\sf R}}^N$.

Then, we recall the partition function for a free realtivistic 
fermionic gas.  As for the case of bosons, one can write this
partition function as a determinant \cite{Kapusta,Actor}
\begin{eqnarray}\label{fermiondet}
{\cal Z}_{Fermions} & = & \left.{\det}_D(i\slash\!\!\!\!D
-M)\right|_{A}
\nonumber\\ & = & \left[\left.\det(-D^2+M^2)\right|_{A}\right]^{+1}\;, 
\end{eqnarray}

\noindent where $\det_D$ means the calculation over
Dirac indices and the subscript $A$ means that the eigenvalues of
each operator are computed with antiperiodic boundary conditions.
Hence, for the operator $-D^2+M^2$, the eigenvalues are given by 
(\ref{eigenvalues}), but now the Matsubara frequencies are 
$\omega_n=(2\pi/\beta)(n+1/2)$ with $n\in {\sf Z}\!\! {\sf Z}$.

Now, with the aid of  a convenient choice of parameters \cite{BBF} 
we are going to use an approach which describe both
the bosonic and fermionic partition functions in an unified way.
Consider then the generalized determinant
\begin{equation}\label{det}
{\cal Z}=\left[\left.\det(-D^2+M^2)\right|_{\theta}\right]^{\sigma}, 
\end{equation}

\noindent  where the subscript $\theta$ means that the quasiperiodic 
finite temperature condition, {\sl i.e.},
\begin{equation}\label{condition}
\psi(0;\vec x)=e^{i\theta}\psi(\beta ;\vec x), 
\end{equation}

\noindent is assumed and we introduced a new parameter $\sigma$ to be
able to reproduce correctly the bosonic and fermionic particular
cases. Observe that when $\theta=0$ and $\sigma=-1$ we obtain the
bosonic partition function, while for $\theta=\pi$ and $\sigma=+1$ we
have the fermionic one.  The eigenvalues in this case are
\begin{equation}\label{eigenvalues2}
\lambda_{n k\theta}= \left[{2n\pi \over \beta} +i\mu + {\theta\over
\beta}\right]^2 +\vec k^2+M^2\;. 
\end {equation}

\noindent The determinant (\ref{det}) is a generalization of its
quantum mechanical $(0+1)$ dimensional version (with $\mu=0$) which
has been calculated using Green functions \cite{BFD} and the zeta
function method \cite{BF}. 

Notice that the interpolating parameter $\theta$ plays the role of the 
zeroth component of a topological (temperature-dependent)
gauge field $A_\mu=(A_0(\beta),\vec 0)$, since the $\theta$-boundary
condition implied that
$\omega_n\rightarrow \omega_n +\theta/\beta$, which can be viewed as a
shift in the time derivative operator, $\partial_0\rightarrow\partial_0
+i\theta/\beta$. The discussion of the bosonic (\ref{bosondet}) and
fermionic (\ref{fermiondet}) determinants with a constant topological
gauge field, but without relating it to boundary conditions or
interpolating statistics, has been presented before \cite{Actor}. It is
important to mention that the parameter $\theta$ or equivalently the
$A_0(\beta)$ field can not be gauged away because of the nontrivial boundary
conditions introduced by the time compactification to the interval
$[0,\beta]$. An interesting point that has not been observed before is
that the coupling between bosons with the topological field $A_0(\beta)$
suppresses the Bose-Einstein condensation, unless for very specific
values of $\theta$, as we are going to show.

It can be shown that the free energy corresponding to the partition
function (\ref{det}), is given by \cite{BBF}:
\begin{eqnarray} 
\Omega(\beta,\mu)&\equiv& - {1\over \beta} \ln {\cal Z} 
= - {\sigma \over \beta}  \ln \det {\cal A}\nonumber\\ &=& 4 \sigma V
\left({ M\over 2\pi \beta}\right)^{N+1\over 2} 
\sum_{n=1}^{+\infty}\cos(n\theta)\cosh(n\beta\mu) 
\left({1 \over n}\right)^{N+1\over 2}
K_{{1 \over 2} (N+1)}(n\beta M)\;, 
\label{Omega}
\end{eqnarray}

\noindent where $K_\nu(x)$ is the modified Bessel function of the second
kind. Note that in the above formula, only the real part of the free
energy $\Omega(\beta,\mu)$ is expressed, according to the prescription
of introducing the chemical potential as an imaginary time-component
gauge potential \cite{Kapusta}. If we particularize the parameters
$\sigma$ and $\theta$ to the bosonic ($\sigma=-1$ and $\theta=0$) and
fermionic ($\sigma=+1$ and $\theta=\pi$) cases we shall find precisely
the results known in the literature \cite{HW,Greiner et al,Actor}. In a
recent work \cite{BBF}, we have shown that this partition function with
arbitrary $\theta$ leads, in the nonrelativistic limit, to a second
virial coefficient which is equivalent to the well known result of
Arovas {\sl et. al} \cite{Arovas}. We have also calculated higher order
virial coefficients showing good agreement with anyon perturbation
theory and numerical calculations \cite{BBF2}. There, we have shown
that, for statistical consistency we have to impose that the parameters
$\theta$ and $\sigma$ are related by
\begin{equation}\label{sigma}
\sigma\equiv \sigma(\theta)=-\cos^{-1}\theta.
\end{equation}


\noindent
To see explicitly when the condensation occurs, let us rewrite
the free energy (\ref{Omega}) as \cite{Actor}
\begin{eqnarray}
\Omega(\beta,\mu)
&=&-{TV\sigma(\theta) \over (2\pi)^N} \int d^N k \;\; \Re
\left\{ \ln\left[1-e^{i\theta} e^{-\beta(\omega-\mu)}\right] +
(\mu\rightarrow -\mu) \right\} \nonumber \\
=&-&{TV\sigma(\theta)\over 2 (2\pi)^N} \int d^N k 
\left\{ \ln\left[1+e^{-2\beta(\omega-\mu)}
-2\cos\theta\, e^{-\beta(\omega-\mu)}\right]
+ (\mu\rightarrow -\mu)
\right\} \label{Omegaprime}
\end{eqnarray}

\noindent where $\omega=\sqrt{\vec k^2 + M^2}$. 
Further, the charge density is given by:
\begin{equation}\label{rho}
\rho\equiv {1\over \beta V}{\partial\over\partial\mu} \ln {\cal Z} 
= -{1\over V}\left({\partial \Omega\over\partial\mu}\right)_{V,T}.
\end{equation}

\noindent Remind that it goes to infinity when particles condense.
From this charge density we can write the general $\theta$-dependent 
distribution function
\begin{eqnarray}
f_\theta(\beta,\pm\mu)
&=&\sigma(\theta) \, 
\Re \left[{1\over 1-e^{-i\theta}e^{\beta(\omega\mp\mu)}}\right] 
\nonumber \\
&=&-{1\over 2\cos\theta}\left[{\cos\theta
-e^{-\beta(\omega\mp\mu)}\over
 \cosh\beta(\omega\mp\mu)-\cos\theta}\right].\label{f_theta}
\end{eqnarray}

\noindent The condensation condition comes from the divergence in the free
energy or equivalently in the corresponding charge density which is also
evident in the above distribution function.  As the condensation is
related to the zero momenta ($\vec k=\vec 0$)  state, this implies that
$\omega\rightarrow M$ and in the general case, with arbitrary $\theta$ we
have that this condition is given by \begin{equation}
1+e^{-2\beta(M-\mu)}-2\cos\theta\; e^{-\beta(M-\mu)}=0\;, \end{equation}

\noindent or simply 
\begin{equation}
\cosh\beta(\mu -M)=\cos\theta\;,
\end{equation}

\noindent which can only be satisfied for finite temperatures if
$\theta=2n\pi$ (with $n$ integer) and $\mu=M$, simultaneously. This means
that the condensation condition in the general case is exactly the same as
in the usual BEC. 

Then, we conclude that equation (\ref{Omegaprime}), or equivalently
(\ref{Omega}), is well defined\footnote{Note that from
(\ref{Omegaprime}) or (\ref{f_theta}) one can see that there is another
possibility of infinite charge density which comes from the choice
$\theta=(2n+1)\pi/2$, since $\sigma\to \pm\infty$, because of relation
(\ref{sigma}). However, this divergence does not depend on the other
physical parameters of the system and then this sigularity opposed to
the BEC case, seems to be unphysical.} for every value of $\mu$, except
when $\theta=2n\pi$ for which $\mu=M$ gives the well known condensation
condition. So, we can infer that no condensation occurs for any
$\theta\not=2n\pi$ (n integer). However, this does not forbid the
condensation of fermions or $\theta$-fields. 

Consider, for instance, that the free $\theta$-field obeying the
Klein-Gordon equation is now interacting with another topological field
$A_0^\prime$, different from the $A_0$ introduced above to transmutate
the statistics of the original bosonic or fermionic field. In this case,
the parameter defining the statistics and condensation condition for the
$\theta$-field together with the topological interaction can be written
as $\Theta=\theta+\theta^\prime$, where $\theta^\prime$ is associated
with the $A_0^\prime$ field . Then, the statistical properties in this
situation are determined by the parameter $\Theta$. This means that
apart from the usual BEC, fermions coupled to topological fields with
$\theta^\prime=n \pi$ so that the total phase is $\Theta=2n\pi$ may
possibly condense. Analogously, any $\theta$-field coupled to a
topological field with $\theta^\prime=\theta-2n\pi$, such that
the total phase is $\Theta=2n\pi$ may also condense.

To obtain the critical temperatures, densities and dimensions for the
condensate from the above discussion we go back to the free energy
given by Eq. ({\ref{Omega}). Note that this equation 
contains all the relevant information about our
system. However, we are mainly interested in the ultra and
nonrelativistic limiting cases. 


Let us first consider the ultrarelativistic case. In this situation
where the energies involved are much higher than the mass scale (the
mass $M$ of the particles) we should take the limit $\beta M<<1$,
which is easily recognized as a high temperature limit. In this regime 
we can use the following property of the Bessel functions
\begin{equation}
\lim_{x\rightarrow 0} K_\nu (x) = \Gamma(\nu) 2^{\nu-1} x^{-\nu}
\;;\;\;\;\;\;\;\;\;\;\;\;\;\;\;\;(\nu>0)
\end{equation}

\noindent so that the free energy (\ref{Omega}) 
reduces to ($\theta=2n\pi$, $\sigma(\theta)=-1$)
\begin{equation}
\Omega(\beta,\mu)=-{2V\over (\beta \sqrt\pi)^{N+1}}
\Gamma({N+1\over 2})\sum_{n=1}^{+\infty}\cosh(n\beta\mu)
({1 \over n})^{N+1}\;.
\label{highOmega}
\end{equation}

\noindent 
Using now Eq. (\ref{rho}) we find that 
\begin{equation}
\rho(\beta,\mu)={2 \over (\sqrt\pi)^{N+1}} 
\left({1\over \beta}\right)^N \Gamma({N+1\over 2})
\sum_{n=1}^{+\infty}\sinh(n\beta\mu)
({1 \over n})^{N}\;.
\end{equation}

\noindent
As we are dealing with the high temperature limit, $\beta M<<1$, the
condensation condition $\mu\rightarrow M$ also implies 
$\beta \mu<<1$, so that
\begin{equation}\label{highrho}
\rho = {2\mu \over \pi^{(N+1)/2}}\Gamma({N+1\over 2}) 
\zeta(N-1) T^{N-1} ,
\end{equation}

\noindent where $\zeta(s)=\sum_{n=1}^\infty n^{-s}$ 
is the usual Riemann zeta function. Using the fact that the critical
temperature of the condensate is reached when $\mu=M$,
in $N=3$ space dimensions, we have:
\begin{equation}
T_c=\left( {3\rho\over M}\right)^{1/2},
\end{equation}

\noindent which coincides with the ulatrarelativistic 
BEC results known in the literature
[\ccite{HW}-\ccite{KirstenToms}]. 


Now, let us study the situation for the nonrelativistc limit which
here means to take the low temperature limit, $\beta M >>1$. 
Using the asymptotic expansion for the
Bessel function
\begin{equation}\label{asymptotic}
K_\nu (x) \simeq \sqrt{\pi\over 2x} e^{-x},
\end{equation}

\noindent valid when $|x|>>1$ and $-\pi/2<\arg x<\pi/2$, and taking
$\beta M >>1$ in the free energy(\ref{Omega}), 
we have ($\theta=2n\pi$, $\sigma(\theta)=-1$):
\begin{equation}
\Omega(\beta,\mu)=-{2V\over\beta} \left({M \over
2\pi\beta}\right)^{N\over 2}
\sum_{n=1}^{+\infty}\cosh(n\beta\mu) ({1 \over
n})^{1+{N\over 2}}e^{-n\beta M}\;.
\end{equation}

\noindent In this regime the charge density reads
\begin{equation}
\rho(\beta,\mu)= 2 \left({M \over 2\pi\beta}\right)^{N\over 2}
\sum_{n=1}^{+\infty}\sinh(n\beta\mu) ({1 \over
n})^{{N\over 2}}e^{-n\beta M}\;. 
\end{equation}

\noindent Exactly as in the ultrarelativistic case,
here the critical point is reached when $\mu\rightarrow
M$. So, in the nonrelativistic limit, the condesation condition
implies that $\beta\mu >>1$ and then the charge density reduces to 
\begin{equation}
\rho(\beta,\mu)= \left({M \over 2\pi\beta}\right)^{N\over 2}
\sum_{n=1}^{+\infty} ({1 \over n})^{{N\over 2}} e^{n\beta(\mu- M)}.
\end{equation}

\noindent At the critical point $\mu=M$ we get simply
\begin{equation}\label{lowrho}
\rho = \zeta({N\over 2})\left({T_c M\over 2\pi}\right)^{N\over 2}, 
\end{equation}

\noindent which for $N=3$ 
space dimensions leads to the critical temperature 
\begin{equation}
T_c={2\pi\over M} \left({\rho\over \zeta(3/2)}\right)^{2/3},
\end{equation}

\noindent in agreement with the nonrelativistic 
BEC results known in the literature 
[\ccite{Huang}-\ccite{KirstenToms}].

Looking at the Eqs. (\ref{highrho}) and (\ref{lowrho}) for the charge
density in $N$ space dimensions for the ultrarelativistic and
nonrelativistic cases one can see that the condensate is not defined in
two space dimensions for both cases, since the Riemann zeta function
$\zeta (s)$ has a pole at $s=1$. Hence, we find that the condensation
for free massive fields (both in the ultra or nonrelativistic limits)
occurs for space dimensions $N>2$, which is in accordance with the BEC
results for the massive case in the absence of external fields 
[\ccite{Huang}-\ccite{KirstenToms}].


We presented here a relativistic partition function from which we
described both the nonrelativistic and ultrarelativistic condensation.
With our partition function we were also able to describe fermions and
bosons at the same time by varying the parameters $\theta$ and $\sigma$.
From this partition function we found the general condensation condition
of the fields, {\sl i.e.}, they should strictly obey {\sl periodic}
boundary conditions at finite temperature. This condition can be
achieved by any field including fermions but conditioned to the fact
that they must be coupled to topological fields which result in a total
phase $\Theta=2n\pi$. On the other side, bosons coupled to those fields
in such a way that $\Theta \not= 2n\pi$ do not condense.

We also described here a kind of finite temperature 
bosonization (or fermionization) by
including the arbitrary $\theta$ in the partition
function. We have shown that no condensation occurs for fields with
$\theta\not=2n\pi$. However, a fermionic field interacting with a
topological field $A_0=\pi / \beta$ will also condense, so this should 
correspond to a bosonization process. Reversely, adding such a field
to a boson we fermionize it once they will exclude rather than
condense. Note that this picture is valid only at finite 
temperatures since at zero temperature the topological field 
$A_0=\theta/\beta$ vanishes.

Our results also suggest that pairs of fermions can also condense as
they would satisfy the condition $\Theta=2n\pi$, as well as arbitrary
combinations of $\theta$-fields as long as that condition is preserved.
However, to have a clear picture of such situations one needs to
consider some kind of interaction between these fields which was not
considered in our approach. This is presently under investigation and
will be reported elsewhere.

\bigskip

\noindent {\bf Acknowledgments.} H.B.-F. acknowledges R. Jackiw 
for reading a preliminary version of the manuscript and for useful 
suggestions on it, and the hospitality of the 
Center for Theoretical Physics - MIT, 
where part of this work was done. 
The authors H.B.-F. and C.F. were partially supported by CNPq
(Brazilian agency).





\end{document}